\documentclass[english,12pt]{article}
\usepackage{array}
\usepackage{graphicx}
\usepackage{amssymb}
\usepackage{amsmath}
\usepackage{multirow}
\usepackage{prettyref}
\usepackage{babel}
\usepackage{units}
\usepackage[latin1]{inputenc}
\usepackage{amsfonts}
\usepackage{amssymb}
\usepackage{babel}
\usepackage{color}
\usepackage{cite}
\def\@fmsl@sh#1#2#3{\m@th\ooalign{$\hfil#1\mkern#2/\hfil$\crcr$#1#3$}}
 \def\eq#1\en{\begin{equation}#1\end{equation}}
\def\s[#1,#2]{[#1\stackrel{\star}{,}#2]}
\def\sx[#1,#2]{[#1\stackrel{\star_{x}}{,}#2]}

\newcommand{\nc}{\newcommand}
\nc{\beq}{\begin{equation}}
\nc{\eeq}{\end{equation}}
\nc{\beqa}{\begin{eqnarray}}
\nc{\eeqa}{\end{eqnarray}}

\def\bc{\begin{center}}
\def\ec{\end{center}}

\def\to{\rightarrow}

\def\gsim{\mathrel{\mathpalette\atversim>}}

\def\bc{\begin{center}}
\def\ec{\end{center}}

\def\gsim{\mathrel{\rlap{\lower4pt\hbox{\hskip1pt$\sim$}}

    \raise1pt\hbox{$>$}}}       

\def\gsim{\mathrel{\rlap{\lower4pt\hbox{\hskip1pt$\sim$}}
    \raise1pt\hbox{$>$}}}       

\begin{document}
\makeatletter
\def\fmslash{\@ifnextchar[{\fmsl@sh}{\fmsl@sh[0mu]}}
\def\fmsl@sh[#1]#2{%
  \mathchoice
    {\@fmsl@sh\displaystyle{#1}{#2}}%
    {\@fmsl@sh\textstyle{#1}{#2}}%
    {\@fmsl@sh\scriptstyle{#1}{#2}}%
    {\@fmsl@sh\scriptscriptstyle{#1}{#2}}}
\def\@fmsl@sh#1#2#3{\m@th\ooalign{$\hfil#1\mkern#2/\hfil$\crcr$#1#3$}}
\makeatother

\thispagestyle{empty}
\begin{titlepage}
\boldmath
\begin{center}
  \Large {\bf Transformation properties and entanglement of relativistic qubits under space-time and gauge transformations}
    \end{center}
\unboldmath
\vspace{0.2cm}
\begin{center}
{  {\large Xavier Calmet$^*$}\footnote{x.calmet@sussex.ac.uk} {\large and} {\large Jacob Dunningham\footnote{j.dunningham@sussex.ac.uk}} }
 \end{center}
\begin{center}
{\sl Department of Physics $\&$ Astronomy, 
University of Sussex, Brighton, BN1 9QH, United Kingdom 
}
\end{center}
\vspace{5cm}
\begin{abstract}
\noindent
We revisit the properties of qubits under Lorentz transformations and, by considering Lorentz invariant quantum states in the Heisenberg formulation, clarify some misleading notation that has appeared in the literature on relativistic quantum information theory. We then use this formulation to consider the transformation properties of qubits and density matrices under space-time and gauge transformations. Finally we use our results to understand the behaviour of entanglement between different partitions of quantum systems. Our approach not only clarifies the notation, but provides a more intuitive and simple way of gaining insight into the behaviour of relativistic qubits. In particular, it allows us to greatly generalize the results in the current literature as well as substantially simplifying the calculations that are needed.

\end{abstract}  
\end{titlepage}



\newpage

\section{Introduction}

Entanglement is one of the most fundamental phenomena in quantum physics and underpins the rapidly growing research areas of quantum information and quantum technology \cite{nielsen_chuang, Dowling}. Motivated by its fundamental importance as well as promises of new applications, the theory of entanglement has received a lot of attention over the past three decades. To date, these studies have focused almost exclusively on the realm of nonrelativistic quantum mechanics. However, more recently attention has turned to relativistic treatments, both as a more complete description of the underlying physics and to understand any new features or behaviours that this more general framework reveals.

A growing body of research on different quantum systems has uncovered a wealth of results about how relativity affects entanglement \cite{czachor_einstein-podolsky-rosen-bohm_1997, gingrich, Peres:2002wx, dunningham2009,peres_quantum_2002, alsing_lorentz_2002, ahn2003a, ahn2003b, czachor_relativistic_2003, moon_relativistic_2004, lee_quantum_2004, czachor_comment_2005, lamata_relativity_2006, alsing_entanglement_2006, chakrabarti_entangled_2009, fuentes_entanglement_2010, Friis:2009va}. A clearer picture is emerging of relativistic quantum information theory and how qubits are transformed and entanglement is affected in different frames.
However some misleading notation has propagated in the literature that we want to address. Clarifying the notation helps us understand the physical situation and gives much better intuition into the behaviour of these systems.
In this paper, we begin by reviewing the transformations of qubits under Lorentz and gauge transformations. We then extend this to density matrices and finally consider the transformation of entanglement in different rest frames. Our approach reveals a more intuitive and simple way of understanding the behaviour of different entanglement partitions under Lorentz boosts. This new approach is checked against existing work \cite{Friis:2009va} and shown to be consistent with it. However, significantly, we show that it extends far beyond it. Whereas previous work considered limited and somewhat artificial quantum states with a particular geometry, we are able to consider very general systems with arbitrary boost geometries, general momenta, multiple qubits and multi-level quantum systems (qudits). The added advantage of our approach is that, despite being more general, the calculations are very much simpler. This formalism could also be extended to different measures of entanglement or other interesting observables that could give further insight into relativistic quantum information theory.

\section{Lorentz and gauge transformations}

In the relativistic quantum information theory literature, a relativistic ket for a single particle is typically defined by \cite{gingrich,Peres:2002wx}
\begin{eqnarray} \label{defKet}
|\Psi \rangle =\sum_\sigma \int_{-\infty}^\infty d\mu(p) \psi_\sigma |\sigma,p \rangle,
\end{eqnarray}
where $\sigma$ and $p$ are respectively the spin and momentum, $\psi_\sigma=\langle \sigma,p|\Psi \rangle$ is the wave function, and $|\sigma,p \rangle$ is a one-particle state.
The Lorentz invariant measure is given by \cite{Weinberg:1995mt}

\begin{eqnarray}
d\mu(p) =\frac{d^4 p}{(2\pi)^3} \delta(p^2-m^2) \theta(p^0) \to \frac{d^3 p}{(2\pi)^3 2 p^0},
\end{eqnarray}
where $\theta(p^0)$ is the Heaviside function and the last step holds because, as we shall see shortly, $\psi_\sigma |\sigma,p \rangle$ is a Lorentz invariant function $f (p)$ of the four-momentum $p_\mu$ of a particle of mass $m$ with positive energy $p^0>0$. Let us first observe that 
$|\Psi \rangle$ does not carry any space-time index: on the right hand side of (\ref{defKet}) one sums over the spin and integrates over the four-momentum. It is thus a Lorentz scalar, as expected when using the Heisenberg picture. It is worth emphasizing that the definition given by (\ref{defKet}) only makes sense if one uses the Heisenberg picture.  The ket $|\Psi \rangle$ is a Lorentz scalar, it does not carry any Lorentz or space-time index and it is thus invariant under Lorentz transformations. 

In the literature (see e.g. \cite{gingrich,Peres:2002wx, dunningham2009}) one often finds the misleading notation $|\Psi \rangle^\prime=U(\Lambda) |\Psi \rangle$ where $U(\Lambda)$ is a Lorentz transformation. This ket cannot be transformed in this way because it is a Lorentz scalar and thus invariant. On the other hand the wave-function $\psi_\sigma$ and the one-particle state $|\sigma,p \rangle$ are Lorentz transforming quantities. For a spin $1/2$ particle, one has
\begin{eqnarray}
U(\Lambda) |\sigma,p \rangle = \sum_\lambda D_{\lambda,\sigma}[W(\Lambda,p)]|\lambda, \Lambda p \rangle
\end{eqnarray}
under a Lorentz transformation. The transformation of the wave function can be found from the invariance of the ket $|\Psi \rangle$ as follows
\begin{eqnarray}
\psi_\sigma^\prime(p^\prime)=(\langle \sigma,p|)^\prime |\Psi \rangle
&=&\sum_\lambda D^{\dagger}_{\lambda,\sigma} [W(\Lambda,p)] 
\langle \lambda, \Lambda p|\Psi \rangle \\ \nonumber 
&=&\sum_\lambda D^{-1}_{\lambda,\sigma} [W(\Lambda,p)] 
\langle \lambda, \Lambda p|\Psi \rangle \\ \nonumber 
&=&
\sum_\lambda D_{\lambda,\sigma} [W(\Lambda^{-1},p)] 
\langle\lambda, \Lambda p|\Psi \rangle \\ \nonumber 
&=&
\sum_\lambda D_{\lambda,\sigma} [W(\Lambda^{-1},p)] 
\psi_\lambda(\Lambda p).
\end{eqnarray}
It is straightforward to generalize this to a two-particle system:
\begin{eqnarray}
\psi_{\sigma_1^\prime,\sigma_2^\prime}(p^\prime,q^\prime)= \sum_{\lambda_1,\lambda_2} D_{\lambda_1,\sigma_1}[W(\Lambda^{-1},p)]    D_{\lambda_2,\sigma_2} [W(\Lambda^{-1},q)] 
\psi_{\lambda_1,\lambda_2}(\Lambda p,\Lambda q).
\end{eqnarray}

We can see that the bras and the kets, besides being Lorentz invariant, are also gauge invariant \cite{Calmet:2012pn}. If we consider quantum electrodynamics, i.e. a local gauge transformation, the wave-function of the spinor field transforms as $\psi^\prime(x)=\exp ( i \alpha(x)) \psi(x)$ and the gauge potential as $A_\mu^\prime(x)= A_\mu(x) -1/e \partial_\mu \alpha(x)$ where $e$ is the electric charge. The one-particle state $|\sigma,p\rangle$ transforms as $|\sigma,p\rangle^\prime = \exp ( -i \alpha(x)) |\sigma,p\rangle$ and we thus see that the ket $|\Psi \rangle$ is invariant under gauge transformations. 


We now turn our attention to the density matrix. We begin by defining this object and then consider its transformation properties. We follow the presentation in Feynman's book on statistical mechanics \cite{Feynman}.  In particular, it is important to realize in which space it is defined. Let the variable $x$ describe the coordinates of the system and $y$ the rest of the universe. The most general wave function can then be written as $\Psi(x,y)=\sum_i C_i(y) \phi_i(x)$.  

Using Dirac notation we introduce $\{|\phi_i\rangle\}$ which is a complete set of vectors in the vector space describing the system (the complete set is Lorentz invariant in $x$-space) and $\{|\theta_i\rangle\}$ (which is Lorentz invariant in $y$-space) is a complete set for the rest of the universe: $\phi_i(x)=\langle x |\phi_i\rangle$ and $\theta_i(y)=\langle y|\theta_i\rangle$. The most general wave-function is thus 
\begin{eqnarray}
\psi(x,y)=\langle y |\langle x |\psi\rangle= \sum_{i,j} C_{i,j}\langle x |\phi_i\rangle \langle y|\theta_j\rangle
\end{eqnarray}
where $C_{i,j}$ are constant complex numbers.  Under Lorentz transformations, $\psi(x,y)$ transforms as $\psi(x^\prime,y^\prime)=U(\Lambda_x)U(\Lambda_y) \psi(x,y)$. The density matrix is defined by
\begin{eqnarray}
\rho_{i^\prime i} =\sum_j C_{i,j}^\star C_{i^\prime,j}
\end{eqnarray}
$\rho_{i^\prime i}$ are in $\mathbb{C}$ and are thus scalars under Lorentz transformations -- they do not transform. The density matrix operator $\rho$ is defined such that 
\begin{eqnarray} \label{rho}
\rho_{i^\prime i} =  \langle\phi_i^\prime|\rho|\phi_i\rangle
\end{eqnarray}
and since $\rho$ only operates on the system described by $x$,  is it also a Lorentz invariant quantity.

One can also introduce an $x$-representation for the density matrix 
\begin{eqnarray}
\rho(x^\prime,x) \equiv  \langle x^\prime |\rho |x \rangle = \int \psi(x^\prime,y) \psi^\star(x,y) d\mu(y)
\end{eqnarray}
and in momentum space:
\begin{eqnarray}
\rho_{\sigma^\prime(p^\prime),\sigma(p)}(p^\prime,p) \equiv  \langle \sigma^\prime(p^\prime), p^\prime |\rho |  \sigma(p), p \rangle = \int \psi(p^\prime,q) \psi^\star(p,q) d\mu(q).
\end{eqnarray}
It is straightforward to check that under a Lorentz transformation one has $\rho(x^\prime,y^\prime)=U(\Lambda_x)\rho(x,y)U(\Lambda_y)^\dagger$ and $\rho(q^\prime,p^\prime)=U(\Lambda_p) \rho(q,p)U(\Lambda_q)^\dagger$. 

For fermions, the density matrix considered in Eq. (\ref{rho})  can be expressed as
\begin{eqnarray} \label{DM}
\rho=\sum_{\sigma_1(p),\sigma_2(q)} \int \int d\mu(p) d\mu(q)
\rho_{\sigma_1(p),\sigma_2(q)}(p,q) |p,\sigma_1(p) \rangle \langle q,\sigma_2(q)|
\end{eqnarray}
which is a Lorentz scalar. One can obtain a reduced density matrix by considering
\begin{eqnarray}
\int d\mu(r) \langle r| \rho|r \rangle=\sum_{\sigma_1(p),\sigma_2(q)} \int d\mu(r)
\rho_{\sigma_1(p),\sigma_2(q)}(r,r) |\sigma_1 (p)\rangle \langle \sigma_2(q)| \label{oneparticlered}
\end{eqnarray}
where $r$ is a momentum. This reduced density matrix is, however, not Lorentz invariant as $|r \rangle$ does not have well-defined Lorentz transformation properties on its own. This observation agrees with a body of literature that has considered single particles with spin and shown that under Lorentz transformations a partial trace over momentum or spin (and hence the entanglement between momentum and spin) is not invariant\cite{peres_quantum_2002, gingrich, Palge2011a, Palge2012a}. To get a Lorentz invariant, one would have to trace over a complete spinor such as $|\sigma,r \rangle$ which as explained above transforms as $ U(\Lambda) |\sigma,r \rangle = \sum_\lambda D_{\lambda,\sigma}[W(\Lambda,r)]|\lambda, \Lambda r \rangle$, i.e. as a spinor under Lorentz transformations. Note that the transformation involves a sum over the spin index and it is not possible to factorize the spin and momentum transformations since the rotation matrix $D_{\lambda,\sigma}[W(\Lambda,r)]$ depends both on the spin and the momentum.  A Lorentz invariant reduced density matrix can be obtained by calculating 
\begin{eqnarray}
\sum_\lambda \int d\mu(r) \langle r, \lambda(r)| \rho|r, \lambda(r)\rangle=\sum_\lambda \int d\mu(r) \rho_{\lambda,\lambda}(r,r)=1
\end{eqnarray}
which is however trivial.

\section{Entanglement under Lorentz transformations}

In relativity, fundamental quantities are observer independent, however there are many things that may be of interest that are not Lorentz invariant. Entanglement is one such example of particular note for quantum information theory. Entanglement is not a fundamental property of a system but depends on the choice of partition into subsystems and studies have shown that entanglement can be Lorentz invariant for some partitions and not others \cite{Friis:2009va}. We now consider this in the context of the above discussion.

There are different ways of quantifying entanglement \cite{vedral1998}, but for our purposes we define the entanglement between the $i$-th part of the system and the rest of the system to be
\begin{eqnarray}
E_i(\rho) = 1- \mbox{Tr} \rho_i^2,
\end{eqnarray}
where $\rho_i$ is obtained by tracing over all subsystems except the $i$-th. If the $i$-th subsystem is a pure state and not entangled with the rest of the system then the trace over the rest of the system will not affect it, i.e. it will remain in a pure state and hence $E_i(\rho)=0$ as expected. $E_i(\rho)$ will increase as $i$ is increasingly entangled with the rest of the system. We can also sum up the entropies found for different partitions as in \cite{Friis:2009va}
\begin{eqnarray}
E(\rho) = \sum_i (1- \mbox{Tr} \rho_i^2) \label{entropysum}.
\end{eqnarray}
This is not necessarily a Lorentz invariant quantity since, as discussed above, the reduced density matrix $\rho_i$ is not necessarily a Lorentz scalar, but will depend on the choice of spins or momenta that have been traced or summed over. The amount of entanglement between the different states is thus frame dependent. 
A similar observation was made in, e.g. \cite{Friis:2009va}, however as we shall see our formulation gives a simpler and more intuitive way of seeing this. In particular, our approach does not require lengthy calculations.

We begin by extending our considerations to many particle states, which can be done straightforwardly. For example for a two-particle state, the density matrix becomes
\begin{eqnarray} \label{DM2}
\rho&=&\sum_{\sigma_1(p_1),\sigma_2(p_2),\sigma_3(p_3),\sigma_4(p_4)} \int\int\int\int  d\mu(p_1) d\mu(p_2)d\mu(p_3) d\mu(q_4) 
\nonumber \\  && \rho_{\sigma_1(p_1),\sigma_2(p_2),\sigma_3(p_3),\sigma_4(p_4)}(p_1,p_2,p_3,p_4) |p_1,\sigma_1(p_1), p_2,\sigma_2(p_2) \rangle \langle p_3,\sigma_3(p_3), p_4,\sigma_4(p_5)| \label{twoparticlerho}
\end{eqnarray}
which is a Lorentz scalar.  We are now free to perform different contractions of the two-particle density matrix -- these correspond to finding the entanglement entropies for different partitions according to Eq.~(\ref{entropysum}).  It is now clear what partitions will lead to Lorentz invariant entanglements: since the total density matrix $\rho$ is invariant, overall Lorentz invariance will be preserved so long as $\rho$ is sandwiched with a Lorentz invariant term. For example,  the reduced density matrix
 \begin{eqnarray}
\rho_{red}&=& \int  d\mu(p_5) \sum_{\sigma_5(p_5)}\langle p_5,\sigma_5(p_5)| \rho | p_5,\sigma_5(p_5)\rangle \label{particle_partition}
\end{eqnarray}
is Lorentz invariant independently of the choice made for the contraction of the momenta or spin, since the trace is over Lorentz invariant spinors as discussed in Section~2 above.

On the other hand, if we decide to sandwich $\rho$ with a  vector momentum $a$ (as in Eq.~(\ref{oneparticlered})) or spin $\sigma$, we would obtain a reduced density matrix which is frame dependent, e.g. 
\begin{eqnarray}
\rho_{red}&=&  \langle a| \rho |a\rangle \label{atrace}
\end{eqnarray}
is not Lorentz invariant. The same would apply if we integrated over $a$, but did not sum over $\sigma$. We thus have a very simple criterion to check whether a given reduced density matrix is Lorentz invariant or not: the reduction of the density matrix must be done in a covariant way. 

We can compare this general result directly with the specific two-particle case considered in \cite{Friis:2009va}. In that work the authors considered the case of two spin-half particles each with one of two possible delta-function values of momentum $\{p_+,p_-\}$ that are equal and opposite in the $z$-direction. Their initial state was
 \begin{eqnarray}
|\psi_{\rm total}\rangle = (\cos\alpha |p_+,p_-\rangle +\sin\alpha |p_-,p_+\rangle)(\cos\beta|\uparrow \downarrow\rangle\rangle + \sin\beta|\downarrow \uparrow\rangle\rangle),   \label{friis_state}
\end{eqnarray}
where $\alpha$ and $\beta$ are real numbers and the kets respectively represent the momentum of particles 1 and 2 and the spins of particles 1 and 2. This can be considered as a four qubit state with two spin and two momentum qubits. We note that this is a very specific (and somewhat artificial) state with spin-1/2 particles in a particular initial state and momentum delta functions in the $z$-direction; we shall see shortly how our formulation can handle general situations.  
They then subjected the overall state (\ref{friis_state}) to a specific Lorentz transformation in the $x$ direction and performed detailed calculations of the entanglement of this transformed state using (\ref{entropysum}) for different partitions to determine, among other things, which partitions correspond to Lorentz invariant entanglements. The partitions they considered were: 1) any one of the qubits with the other three; 2) the two spin qubits and the two momentum qubits; 3) the particle-particle partition, i.e. the momentum and spin of one particle and the momentum and spin of the other.

We can easily analyse these cases without any need for a calculation. The density matrix is 
\begin{eqnarray}
\rho = |\psi_{\rm total}\rangle \langle \psi_{\rm total} |,
\end{eqnarray}
and can use this as the density matrix in our formulation above. For partition 1, we need to trace over any one of the spin or momentum qubits. As argued in Eq.~(\ref{atrace}) and the text below it, this leads to a reduced density matrix that is not Lorentz invariant and hence nor is the entanglement. Similarly for partition 2, we need to trace over both spins or both momenta. For the spins, the reduced density matrix is
 \begin{eqnarray}
\rho_{\rm mom}&=&  \sum_{\sigma_5(p_5), \sigma_6(p_6)}\langle \sigma_5(p_5), \sigma_6(p_6)| \rho |\sigma_5(p_5), \sigma_6(p_6)\rangle, \label{spin_momentum}
\end{eqnarray}
which is not Lorentz invariant. The final partition considered in \cite{Friis:2009va} is partition 3, which is the particle-particle partition and the reduced density matrix for this is given by Eq.~(\ref{particle_partition}) above, which we have shown to be Lorentz invariant. In fact, we can see that this is the only Lorentz invariant partition. 

All of these results are consistent with the findings in \cite{Friis:2009va}. However there they considered a specific state and a specific boost and carried out detailed calculations of the entanglement for each different partition. Only by looking at the results of these calculations were they able to comment on whether different partitions had Lorentz invariant entanglements. By contrast, our formulation is much simpler and can be applied to general states and boosts since it does not rely on the details of the state or geometry. It gives a clear intuition for which partitions are invariant and why and this without performing any calculation. It can also easily be extended to other partitions not considered in  \cite{Friis:2009va} such as a trace over the momentum of one particle and the spin of the other. The reduced density matrix in this case is
 \begin{eqnarray}
\rho_{red}&=& \int  d\mu(p_6) \sum_{\sigma_5(p_5)}\langle p_6,\sigma_5(p_5)| \rho | p_6,\sigma_5(p_5)\rangle,
\end{eqnarray}
which can be seen to be not Lorentz invariant, without any direct calculation. If we were to follow \cite{Friis:2009va}, this conclusion would require a lengthy calculation and, even then, would only be true for the particular initial state and particular boost geometry chosen. Similarly, we could easily extend our method to consider more than two qubits, states with $d$ spin levels (qudits), and general momentum wave packets (i.e. not delta functions). Analyzing these cases using previous methods would be very cumbersome, but we are able to immediately see that it is again just the particle partition that is Lorentz invariant. This illustrates the power and versatility of this formalism.

By considering quantum states in the Heisenberg formulation, we have not only been able to clarify some misleading notation and inaccurate statements, but also provide a simple and useful method for understanding the behaviour of relativistic qubits. This treatment is a good example of when the right notation can be more than simply a convenient way of keeping track of a calculation, but can aid physical insight and understanding. Our formulation enables a much simpler way of drawing general conclusions about the entanglement of quantum systems and could be a useful tool in furthering our understanding of relativistic quantum information theory.

\section{Acknowledgements}
This work was supported in part  by the Science and Technology Facilities Council (Grant number  ST/L000504/1).


\bigskip{}

\end{document}